# New limits on the violation of the Strong Equivalence Principle in strong field regimes

N. Wex*

Max-Planck-Society, Research Unit "Theory of Gravitation", University of Jena, Max-Wien-Platz 1, 07743 Jena, Germany

**Abstract.** The Strong Equivalence Principle (SEP) demands, besides the validity of the Einstein Equivalence Principle, that all self-gravitating bodies feel the same acceleration in an external gravitational field. It has been found that metric theories of gravity other that than general relativity typically predict a violation of the SEP. In case of the Earth-Moon system (weak field system) this violation is called the Nordtvedt effect.

It has been shown by Damour and Schäfer, that small-eccentricity long-orbital-period binary pulsars with a white dwarf companion provide excellent conditions to test the SEP in strong field regimes.

Based on newly discovered binary pulsars this paper investigates a possible violation of the SEP in strong field regimes. New limits with an improved confidence level are presented.

The results of this paper lead to constrains on the combination $\varepsilon/2+\zeta$ of the only two (post)$^2$-Newtonian parameters $\varepsilon$ and $\zeta$ that arise from the (post)$^2$-Newtonian approximation of the tensor-multi-scalar theory of Damour and Esposito-Farèse.

## 1. Introduction

When dealing with relativistic theories of gravity one is confronted with three types of equivalence principles (see Will 1993):

– the Weak Equivalence Principle (WEP),
– the Einstein Equivalence Principle (EEP),
– and the Strong Equivalence Principle (SEP).

The WEP, which goes back to Galileo and Newton, states the universality of free fall for all neutral test masses. In a more geometrical view the WEP states that all test masses move along geodesics in spacetime. Test masses are understood to be bodies with negligible self-energy and therefore with negligible contribution to spacetime curvature. The WEP has been verified with a fractional precision of better than $10^{-11}$ (Roll et al. 1964, Braginsky & Panov 1971, Su et al. 1994).

The EEP demands, besides the validity of the WEP, that in local Lorentz frames the non-gravitational laws of physics are those of special relativity. The EEP implies that spacetime has to be curved and thus is the basic ingredient of any metric theory of gravity.

The SEP states, besides the validity of the EEP, the "universality of free fall for self-gravitating bodies". One has to be careful with the notion of a freely falling self-gravitating bodies in an external gravitational field. There is no rigorous definition for the SEP in relativistic theories of gravity. Because of non-linearity the split of the metric field into an external and a local part can only be approximate. For a discussion of the SEP within a slow-motion weak-field approximation see Ashby & Bertotti (1984) and Bertotti & Grishchuk (1990). For metric theories of gravity, other than general relativity, it has been found that they typically introduce auxiliary gravitational fields (e.g. scalar fields) and thus predict a violation of the SEP (see e.g. Will 1992, Will 1993).

In 1968 Nordtvedt (1968a,1968b) proposed to test a possible violation of the SEP through the analysis of lunar laser ranging (LLR) data. In case of a violation of the SEP, the external field of the sun would cause a "polarization" of the lunar orbit since the Earth and the Moon show different fractions of self-energy. A violation of the SEP implies the that the ratio of passive gravitational mass $m_g$ and inertial mass $m_i$ differs from one by a function of the gravitational self-energy $E^{\text{grav}}$ of the body:

$$\begin{aligned}\frac{m_g}{m_i} &\equiv 1 + \Delta[E^{\text{grav}}] \\ &= 1 + \eta\left(\frac{E^{\text{grav}}}{m_i c^2}\right) + \eta'\left(\frac{E^{\text{grav}}}{m_i c^2}\right)^2 + \ldots\end{aligned} \quad (1)$$

* E-mail: now@gravi.physik.uni-jena.de

Nordtvedt parameter. Present LLR data lead to

$$-0.0016 < \eta < 0.0006 \quad \text{(Dickey et al. 1994)},$$
$$-0.0008 < \eta < 0.0013 \quad \text{(Müller et al. 1995)}. \tag{2}$$

Damour and Schäfer (1991) pointed out that, in view of the smallness of the self-gravity of planetary bodies (e.g. $E^{\text{grav}}/m_i c^2 = -4.6 \times 10^{-10}$ for the Earth), such solar-system tests of the SEP indicate nothing about higher-order gravitational-energy contributions to the ratio $m_g/m_i$.

For binary pulsars the situation is different, since for a neutron star one finds $E^{\text{grav}}/m_i c^2 \sim -0.2$, which takes us into the *strong field regime*. Especially small-eccentricity long-orbital-period binary pulsars with a white-dwarf companion ($E^{\text{grav}}/m_i c^2 \sim -10^{-4}$) constitute an excellent laboratory to test higher-order effects in the violation of the SEP.

Only recently, Damour and Esposito-Farèse (1995) presented a new, field-theory-based framework for discussing and interpreting experimental tests of relativistic gravity, notably at the (post)$^2$-Newtonian order. They use a model in which gravity is mediated by a tensor field together with one or several scalar fields ("tensor-multi-scalar" theories, see Damour & Esposito-Farèse 1992). Damour and Esposito-Farèse come to the conclusion that within their framework the (post)$^2$-Newtonian deviations from general relativity can be fully described by introducing two new parameters, $\varepsilon$ and $\zeta$, beyond the usual (Eddington) parameters $\bar{\beta} \equiv \beta - 1$ and $\bar{\gamma} \equiv \gamma - 1$ of the standard parameterized post–Newtonian (PPN) formalism (see Will 1993). They further find that the parameters $\varepsilon$ and $\zeta$ are practically immeasurable by present and near future solar-system experiments. They conclude that, because of the importance of self-gravity effects in neutron stars, binary-pulsar experiments are an ideal testing ground for the (post)$^2$-Newtonian structure of relativistic gravity, e.g. small-eccentricity long-orbital-period binary pulsars with a white-dwarf companion can be used to set limits on the combination $\varepsilon/2 + \zeta$.

In this paper we investigate newly discovered binary pulsars (Taylor & al. 1993, Taylor et al. 1995) with regard to a violation of the SEP. In Sect. 2 we display the theoretical background for testing the SEP in small-eccentricity long-orbital-period binary pulsars. Sect. 3 introduces the test systems. In Sect. 4 we present the results and in Sect. 5 we give the conclusions.

## 2. Violation of the SEP and binary motion

In this section we repeat briefly the calculations given by Damour & Schäfer (1991). The masses of pulsar and companion will be denoted by $m_p$ and $m_c$, respectively. The total mass of the binary system is given by $M \equiv m_p + m_c$. The ratio of passive gravitational mass and inertial mass for pulsar and companion will be denoted by $1 + \Delta_p$ and

The orbital period of the binary system will be denoted by $P_b$ and the orbital eccentricity by $e$. The eccentricity vector **e** is a vector directed along the apsidal line towards the periastron.

In case of a violation of the SEP the equations of relative motion $\mathbf{r}(t)$ have the form

$$\ddot{\mathbf{r}} + \mathcal{G} M \frac{\mathbf{r}}{r^3} = \mathbf{a}_R + \Delta \mathbf{g}, \tag{3}$$

where $\mathcal{G}$ denotes the effective gravitational constant for the interaction between $m_p$ and $m_c$, $\mathbf{a}_R$ is the orbital relativistic corrections of post-Newtonian order, and **g** denotes the gravitational acceleration field of the Galaxy. **g** is practically constant in magnitude and direction, so that we can interpret $\Delta \mathbf{g}$ as a *"gravitational Stark effect"* caused by a possible violation of the SEP.[**]

For small-eccentricity long-orbital-period binary pulsars Eq. (3) implies that the *orbital plane is fixed* and that the eccentricity vector $\mathbf{e}(t)$ is given by the following vectorial superposition:

$$\mathbf{e}(t) = \mathbf{e}_\Delta + \mathbf{e}_R(t). \tag{4}$$

The constant vector $\mathbf{e}_\Delta$ is directed along the projection of **g** onto the orbital plane ($\mathbf{g}_\perp$),

$$\mathbf{e}_\Delta = \frac{3\Delta}{2\omega_R n a} \mathbf{g}_\perp. \tag{5}$$

$\omega_R$ is the average angular velocity of the relativistic advance of the periastron, $n \equiv 2\pi/P_b$, and $a$ is the semi-major axis of the relative motion. $\mathbf{e}_\Delta$ represents a constant, **g**-induced, polarization of the orbit. $\mathbf{e}_R(t)$ is a vector of constant length which lies in the orbital plane and rotates with angular velocity $\omega_R$, and it represents the usual relativistic advance of the periastron. Figure 1 illustrates the temporal behaviour of the eccentricity vector **e**.

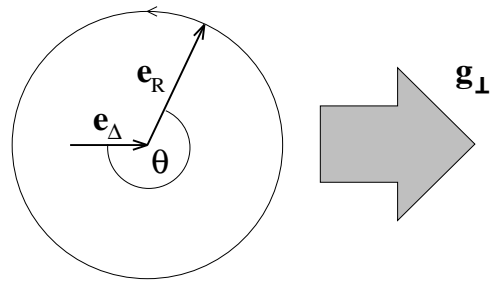

**Fig. 1.** Evolution of the eccentricity vector $\mathbf{e}(t) = \mathbf{e}_\Delta + \mathbf{e}_R(t)$

It is obvious that the limit of validity of Eq. (4) is that $\omega_R$ should be appreciably larger than the angular velocity of reference frame of the binary system.

---
[**] We assume that the limits for a violation of the WEP obtained on Earth apply also for white-dwarf and neutron-star matter.

Allen & Martos (1986) developed a mathematically simple Galactic mass model that gives a good fit to the observed Galactic accelerations. This model will be used here to determine **g** at any given point in the Galaxy. In the Galactic plane we find then for circular orbits the following approximation:

$$\tilde{\omega} \approx \frac{0\overset{''}{.}04/\text{yr}}{\rho[\text{kpc}]} \qquad \text{for} \quad \rho = 1 \ldots 10 \text{ kpc}, \tag{6}$$

where $\rho$ denotes the distance to the Galactic center. Most of the binary pulsars considered below are rather close to the Galactic plane, i.e. Eq. (6) can be used safely to estimate the ratio $\omega_R/\tilde{\omega}$ for most of our test systems. For $\omega_R$ one finds

$$\omega_R = \frac{3\mathcal{F}\mathcal{G}Mn}{c^2 a(1-e^2)}, \tag{7}$$

where in general relativity the factor $\mathcal{F}$ is unity and $\mathcal{G} \equiv G_{\text{Newton}} \equiv G$. For small-eccentricity binary pulsars in the solar vicinity ($\rho \approx 8$ kpc) one finds ($\mathcal{F} \equiv 1$, $\mathcal{G} \equiv G$):

$$\omega_R/\tilde{\omega} \approx 1.4 \left(\frac{P_b}{10^3 \text{days}}\right)^{-5/3} \left(\frac{M}{M_\odot}\right)^{2/3}. \tag{8}$$

Thus only for pulsars with $P_b \gtrsim 10^3$ days we expect that we cannot use the simple solution Eq. (4).

For a small-eccentricity binary pulsar with observed eccentricity $e$ and $\omega_R$ being appreciably larger than $\tilde{\omega}$, the inequality

$$e_\Delta \leq e\,\xi(\theta), \quad \xi(\theta) = \begin{cases} 1/\sin\theta & \text{for} \quad \theta \in [0, \tfrac{1}{2}\pi) \\ 1 & \text{for} \quad \theta \in [\tfrac{1}{2}\pi, \tfrac{3}{2}\pi] \\ -1/\sin\theta & \text{for} \quad \theta \in (\tfrac{3}{2}\pi, 2\pi) \end{cases} \tag{9}$$

holds (see Fig. 1). Using the orbital inclination, $i$, the longitude of the ascending node, $\Omega$, and the angle between the direction of sight and **g**, $\lambda$, one obtains from Eq. (9) the following limit on the violation of the SEP:

$$|\Delta| \leq \frac{e}{\hat{e}} f_{i,\lambda}(\theta, \Omega), \tag{10}$$

where

$$\hat{e} = \frac{gc^2}{2\mathcal{F}\mathcal{G}Mn^2} \tag{11}$$

and

$$f_{i,\lambda}(\theta, \Omega) = \frac{\xi(\theta)}{[1 - (\cos i \cos \lambda + \sin i \sin \lambda \sin \Omega)^2]^{1/2}}. \tag{12}$$

On the right-hand side of Eq. (10) we have eight quantities, six of which can be deduced from observations.

the binary system one gets with rather high precision directly from pulsar timing.
- $g$ and $\lambda$ can be calculated using the Galactic-acceleration model of Allen & Martos (1986). These two quantities depend on the location of the binary pulsar in the Galaxy and thus on the direction of sight and on the distance Earth–pulsar. The distance of a pulsar is calculated from the observed dispersion in the signal, based on measurements of the density of the interstellar distribution of free electrons (Taylor & Cordes 1993). The error in distance is typically of the order of 25%. The error in the direction of sight is extremely small.
- The inclination $i$ and the total mass $M$ can be determined if one knows the mass of the pulsar $m_p$ and the mass of the companion $m_c$:

$$M = m_p + m_c, \qquad \sin i = \frac{M}{m_c} \frac{c\, x\, n^{2/3}}{(\mathcal{G}M)^{1/3}} \tag{13}$$

where $x = a_p \sin i/c$ is the (observable) projected semi major-axis of the pulsar orbit.

The analysis of neutron-star neutron-star binary pulsars implies that with 95% confidence the lower bound for neutron-star masses is between $1.01 M_\odot$ and $1.34 M_\odot$ and the upper bound is between $1.43 M_\odot$ and $1.64 M_\odot$ (Finn 1994). Since during the mass transfer a considerable amount of mass could have accreted onto the neutron star (Bhattacharya & van den Heuvel 1991) the upper limit for our test systems could exceed $1.64 M_\odot$. The combined investigation of X-ray binaries and radio-binary pulsars (Thorsett et al. 1993, van Kerkwijk et al. 1995) suggests $1.8 M_\odot$ as an (conservative) upper limit for the mass of a pulsar. Thus we will use $1.0 M_\odot \leq m_p \leq 1.8 M_\odot$.

The mass of the companion white dwarf is limited by theoretical studies of stellar evolution in binary stars (Joss et al. 1987, Savonije 1987, Rappaport et al. 1995). For small-eccentricity binary pulsars $m_c$ is given as a function of the orbital period $P_b$.

- In case of the small-eccentricity binary pulsars PSR J1713+0747 and PSR B1855+09 (will not be used below) it was possible to measure the Shapiro delay caused by the companion. By this we get further limits on $m_p$ and $m_c$ (Camilo et al. 1994, Kaspi et al. 1994).
- The angles $\theta$ and $\Omega$ are not observable. We shall treat them as independent random variables, uniformly distributed between 0 and $2\pi$. Thus, one can only give limits on $|\Delta|$ with a certain probability by excluding those areas of the $\theta$-$\Omega$ parameter space where $f_{i,\lambda}$ is larger than some given value.

## 3. The test systems

To be able to apply Eq. (10) we need binary pulsars which show the following properties:

nary system with a large value for $P_b^2/e$. As a selecting criterion for useful test systems one finds $P_b^2/e \geq 10^7$ days$^2$. (Including binary systems with $P_b^2/e < 10^7$ days$^2$ would not change the results of this paper, since they put only weak limits ($\gtrsim$ few %) on $|\Delta|$.)

- The binary pulsar must be so old (as member of a clean binary system) that $\mathbf{e}_R$ has had the time to make many revolutions:

$$t_{\rm age} \gg \frac{2\pi}{\omega_R} \approx 2 \times 10^8 {\rm yr} \left(\frac{P_b}{10^3 {\rm days}}\right)^{5/3} \left(\frac{M}{M_\odot}\right)^{-2/3}. \quad (14)$$

- The angular velocity of the periastron advance, $\omega_R$, has to be appreciably lager than the Galactic rotation (see Eq. (8)).

In the latest pulsar catalog (Taylor et al. 1993, Taylor et al. 1995) we find more than 40 pulsars which are member of a binary system. Table 1 shows all small-eccentricity binary pulsars with $P_b^2/e > 10^7$ days$^2$.

**Table 1.** List of test systems. The number given for the eccentricity is the measured eccentricity plus the $2\sigma$-error. The values for $\hat{e}$ were obtained by using $m_p = 1.8 M_\odot$ and the upper limit for $m_c$.

| Pulsar | $e \leq$ | $P_b$ [days] | $P_b^2/e \geq$ [$10^7$ days$^2$] | $\hat{e} \geq$ |
|---|---|---|---|---|
| B0820+02 | 0.012 | 1232.5 | 12.8 | 6.9 |
| J1455-3330 [I] | 0.00017 | 76.2 | 3.4 | 0.038 |
| J1640+2224 [II] | 0.00080 | 175.5 | 3.9 | 0.21 |
| J1643-1224 [I] | 0.00051 | 147.0 | 4.2 | 0.25 |
| J1713+0747 | 0.000075 | 67.8 | 6.1 | 0.031 |
| B1800−27 | 0.00051 | 406.8 | 31.9 | 1.4 |
| B1953+29 | 0.00033 | 117.3 | 4.2 | 0.082 |
| J2019+2425 | 0.00011 | 76.5 | 5.3 | 0.037 |
| J2229+2643 [III] | 0.00026 | 93.0 | 3.4 | 0.053 |

The binary pulsar PSR B0820+02 appears to have a hot, rather young white dwarf companion with an age of $2 \times 10^8$ yr (Koester et al. 1992). Thus Eq. (14) is not satisfied for PSR B0820+02 for reasonable values of $M$. The binary pulsar PSR B1800−27 shows the highest $P_b^2/e$ and consequently would be at present the best test laboratory for a violation of the SEP. The period of the periastron advance is of the order of $3 \times 10^7$ years. Unfortunately the evolutionary history of this pulsar is unclear. Like PSR B0820+02, which shows a similar position in the $P$-$\dot{P}$ diagram (see Fig. 2), this binary system in its present form could be young, i.e. one cannot exclude that mass transfer took place within the past $10^8$ years, which would lead to a violation of Eq. (14). This is supported by the fact that the characteristic age of the pulsar is only $3 \times 10^8$ years, of the last mass transfer. Thus, for the rest of this paper, we shall handle this test system with some reservation.

---

[I] see Lorimer et al. 1995
[II] see Wolszcan 1995
[III] see Camilo 1995

All the other pulsars in Tab. 1 are recycled millisecond binary pulsars with a white dwarf companion. All these systems are well away from the spin-up limit (see Fig. 2) and thus are old clean systems that clearly satisfy Eq. (14).

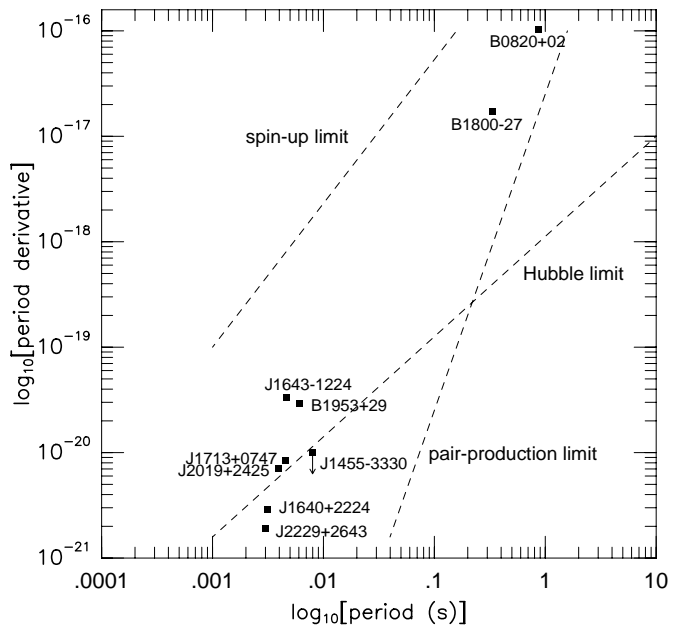

**Fig. 2.** $P$-$\dot{P}$ diagram for the pulsars of Tab. 1

In Eq. (10) we need the total mass $M$ and the orbital inclination $i$ which is equivalent to the knowledge of $m_p$ and $m_c$, see Eq. (13). The restrictions for these quantities are listed in Tab. 2. As mentioned before, in case of PSR

**Table 2.** Limits on the parameters $m_p$, $m_c$, and $i$. For PSR J1713+0747 one gets further limits from the observation of the Shapiro delay caused by the companion

| Pulsar | $m_p$ [$M_\odot$] | $m_c$ [$M_\odot$] | $i$ [deg] |
|---|---|---|---|
| J1455−3330 | 1.0 ... 1.8 | 0.27 ... 0.35 | 0...90 |
| J1640+2224 | 1.0 ... 1.8 | 0.31 ... 0.40 | 0...90 |
| J1643−1224 | 1.0 ... 1.8 | 0.30 ... 0.39 | 0...90 |
| J1713+0747 | 1.2 ... 1.8 | 0.27 ... 0.34 | 74...90 |
| B1800−27 | 1.0 ... 1.8 | 0.36 ... 0.47 | 0...90 |
| B1953+29 | 1.0 ... 1.8 | 0.29 ... 0.37 | 0...90 |
| J2019+2425 | 1.0 ... 1.8 | 0.27 ... 0.35 | 0...90 |
| J2229+2643 | 1.0 ... 1.8 | 0.28 ... 0.36 | 0...90 |

J1713+0747 one was able to measure the Shapiro delay caused by the companion, which can be used to put further limits on $i$, $m_c$, and $m_p$ (Camilo et al. 1994).

Given the orbital period $P_b$, the eccentricity $e$, the total mass $M$, the orbital inclination $i$, and the Galactic acceleration $g$, we can use Eq. (10) to determine the probability for a violation of the SEP, i.e. we calculate the area in the $\theta$-$\Omega$-parameter space where $|\Delta|$ is below a given value. We shall use the approximation $\mathcal{F}\mathcal{G} \simeq G$ in Eq. (10). Using Tab. 2 and assuming an error of 25% in the distance for the binary pulsars we obtain Tab. 3 (see also Wex 1994). Since all these systems represent statistically independent tests for the SEP we can give, besides an individual probability for every binary pulsar, a total probability for the validity of the SEP denoted by $\Sigma$ in Tab. 3. Since the evolutionary history of PSR B1800−27 is uncertain (see previous section), we give two total probabilities, $\Sigma_1$ and $\Sigma_2$, where the results obtained from PSR B1800−27 are excluded in the case of $\Sigma_1$ and included in the case of $\Sigma_2$.

It is important to see that the restrictions on $\Delta$ do not contain any assumptions on the initial eccentricities of the binary systems. In this paper we made a worst case analysis, i.e. we assumed $e_R$ to be such that the largest possible $e_\Delta$ is realised in the observed binary pulsars. If we would have a given probability distribution for $e_R$ we could get a better upper limit on $\Delta$.

In terms of a possible selection effect when choosing the test systems one has to assume that all the binary pulsars with a small $P_b^2/e$ were born like this. A fact that is very well explained by the eccentricity-period relation (Phinney 1992). This re lation explains the existence of an initial eccentricity which is different from zero for all the neutron-star white-dwarf binary pulsars and explains all the observed eccentricities for our test systems.

If we express Eq. (1) in the field-theory based framework of Damour and Esposito-Farèse (1995) and constrain the Nordtvedt parameter $\eta$ by Eqs. (2) then we find for the pulsar

$$\Delta_p = b_p \left(\frac{\varepsilon}{2} + \zeta\right) + \mathcal{O}(c_p^3). \tag{15}$$

Table 3. Probability for the violation of the SEP being below a given $|\Delta|$ for eight test systems

| Pulsar | $|\Delta| \leq$ 1% | $|\Delta| \leq$ 0.5% | $|\Delta| \leq$ 0.4% | $|\Delta| \leq$ 0.3% | $|\Delta| \leq$ 0.2% |
|---|---|---|---|---|---|
| J1455−3330 | 0.82 | 0.32 | 0.00 | 0.00 | 0.00 |
| J1640+2224 | 0.72 | 0.37 | 0.08 | 0.00 | 0.00 |
| J1643−1224 | 0.73 | 0.40 | 0.23 | 0.00 | 0.00 |
| J1713+0747 | 0.92 | 0.82 | 0.77 | 0.61 | 0.00 |
| B1953+29 | 0.74 | 0.44 | 0.00 | 0.00 | 0.00 |
| J2019+2425 | 0.80 | 0.55 | 0.43 | 0.21 | 0.00 |
| J2229+2643 | 0.83 | 0.24 | 0.00 | 0.00 | 0.00 |
| $\Sigma_1$ | 1.00 | 0.99 | 0.91 | 0.69 | 0.00 |
| B1800−27 | 0.95 | 0.92 | 0.90 | 0.87 | 0.80 |
| $\Sigma_2$ | 1.00 | 1.00 | 0.99 | 0.96 | 0.80 |

($\sim -E^{\mathrm{grav}}/mc^2$) of the pulsar. For a medium equation of state one finds

$$b_p \approx 1.03 \, c_p^2, \qquad c_p \approx 0.21 \, m_p/M_\odot. \tag{16}$$

Since $E^{\mathrm{grav}}/m_i c^2 \sim -10^{-4}$ for the white dwarf companion we have $\Delta \simeq \Delta_p$, and thus Tab. 3 gives us limits on $\Delta_p$. Using Tab. 3 and Eq. (15) one finds with 95% confidence for $m_p \approx 1.4 M_\odot$:

$$|\varepsilon/2 + \zeta| \lesssim 5\% \qquad (\text{using } \Sigma_1), \tag{17}$$

$$|\varepsilon/2 + \zeta| \lesssim 3\% \qquad (\text{using } \Sigma_2). \tag{18}$$

In Damour & Esposito-Farèse (1995) one finds with 90% confidence an upper limit of 1.5% for $|\Delta_p|$. This result is solely based on an (obviously less conservative) analysis of the uncertain test system PSR B1800−27 by Arzoumanian (see Damour & Esposito-Farèse 1995).

Bethe and Brown (Brown & Bethe 1994, Bethe & Brown 1995) give an upper limit for neutron-star masses of $1.56 M_\odot$. This limit is supported by the assumption that a low-mass black hole was formed in SN 1987A. If we use this value as an upper limit for $m_p$ in Tab. 2, then we find with 95% confidence respectively 4% and 2.5% as upper limits for $|\varepsilon/2 + \zeta|$.

## 5. Conclusions

In this paper we presented new limits on the violation of the SEP in strong field regimes. The limits where achieved by analyzing long-orbital-period low-eccentricity binary pulsars. We made the conservative assumption that the mass of the pulsar lies between $1 M_\odot$ and $1.8 M_\odot$. The mass of the companion was restricted by results of stellar-evolution scenarios of binary stars. For PSR J1713+0747 we were able to use extra limits on the mass of the pulsar and its companion that came from the observation of the Shapiro delay in the timing data. As a result of the increased number of suitable test systems we can exclude the violation of the SEP being above 0.5% with more than 95% confidence.

We were able to present limits on the tensor-multi-scalar theory of Damour and Esposito-Farèse. For the two parameters of the (post)$^2$-Newtonian approximation of the tensor-multi-scalar theory we found the (conservative) limit $|\varepsilon/2 + \zeta| \lesssim 5\%$ with a confidence of 95%. Including the results of PSR B1800-27 lowered the limit to 3%.

Using $1.56 M_\odot$ as upper limit for the masses of the pulsars, leads to a limit of 4%, respectively 2.5%, for $|\varepsilon/2 + \zeta|$.

*Acknowledgements.* I am grateful to Gerhard Schäfer and Kenneth Nordtvedt for valuable discussions and to Wilhelm Kley for careful reading of the manuscript. I would like to thank Michael Kramer who pointed my attention to the new pulsar catalog of Taylor et al., and Werner Pfau, who brought to my attention the nice work of Allen and Martos.